\documentclass[12pt]{article}
\usepackage{graphicx}
\begin{document}

\centerline{\bf Test of Universality in Anisotropic 3D Ising Model}

\bigskip

M. A. Sumour, Physics Department, Al-Aqsa University, P.O.4051, Gaza, 
Gaza Strip, Palestinian Authority, msumoor@yahoo.com

\medskip
D. Stauffer,Institute for Theoretical Physics, Cologne University, D-50923 K\"oln, Germany, stauffer@thp.uni-koeln.de 

\medskip
M.M. Shabat, Physics Department, Islamic University, P.O.108, Gaza, 
Gaza Strip, Palestinian Authority, shabat@mail.iugaza.edu

\medskip
A.H. El-Astal, Physics Department, Al-Aqsa University, P.O.4051, Gaza, 
Gaza Strip, Palestinian Authority, a\_elastal@yahoo.com

\bigskip
Abstract:

Chen and Dohm predicted theoretically in 2004
that the widely believed universality principle is violated in 
the Ising model on the simple cubic lattice with more than only six nearest 
neighbours. Schulte and Drope by Monte Carlo simulations found such violation, 
but not in the predicted direction. Selke and Shchur tested the square lattice.
Here we check only this universality 
for the susceptibility ratio near the critical point.
For this purpose we study first the standard Ising model on a simple cubic lattice with six nearest neighbours, then with six nearest and twelve next-nearest neighbours, and compare the results with the Chen-Dohm lattice of six nearest 
neighbours and only half of the twelve next-nearest neighbours. We do not 
confirm the violation of universality found by Schulte and Drope in the 
susceptibility ratio.

\bigskip
\begin{figure}[hbt]
\begin{center}
\includegraphics[angle=-90,scale=0.5]{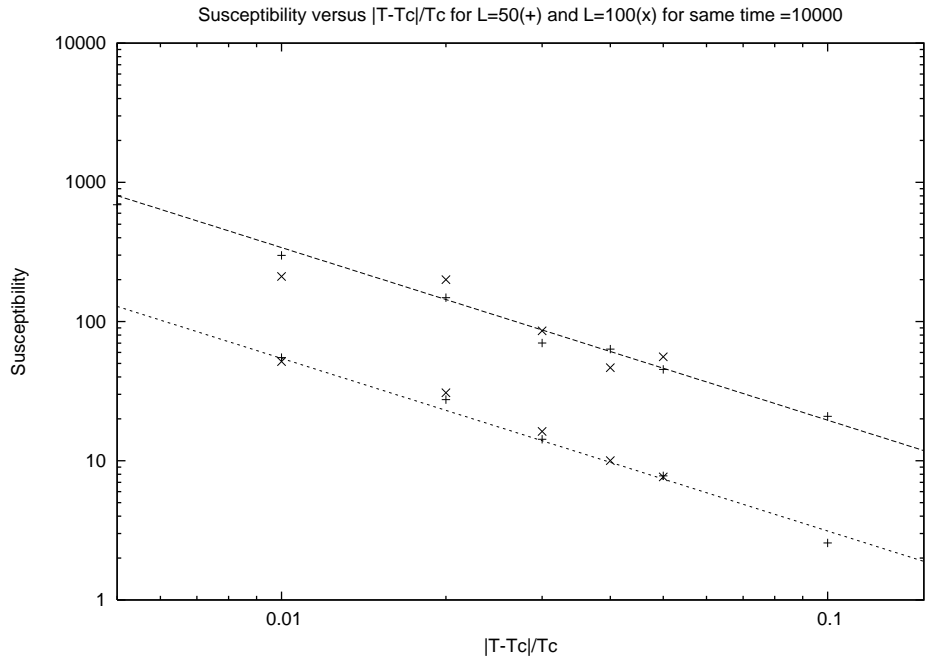}
\end{center}
\caption{
Susceptibility versus temperature difference for $L=50$ and 100 for 6 nearest 
neighbours as log-log plot. The upper data correspond to $T > T_c$, the lower 
to $T < T_c$. Always 1000 iterations were made.
The straight lines in Figs.1,3,4 have the theoretical slope $-1.24$.
}
\end{figure}

\begin{figure}[hbt]
\begin{center}
\includegraphics[angle=-90,scale=0.5]{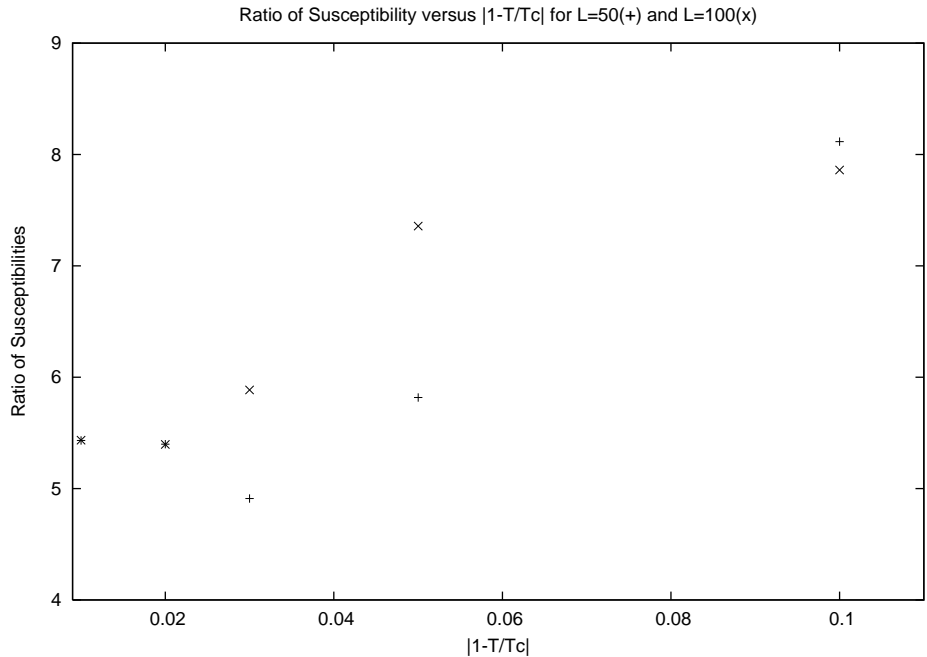}
\end{center}
\caption{
Ratio of susceptibilities above to below $T_c$,  versus $|1-T/T_c|$,
for the two lattices $L=50$ and 100 for 6 nearest neighbours.
}
\end{figure}

\begin{figure}[hbt]
\begin{center}
\includegraphics[angle=-90,scale=0.31]{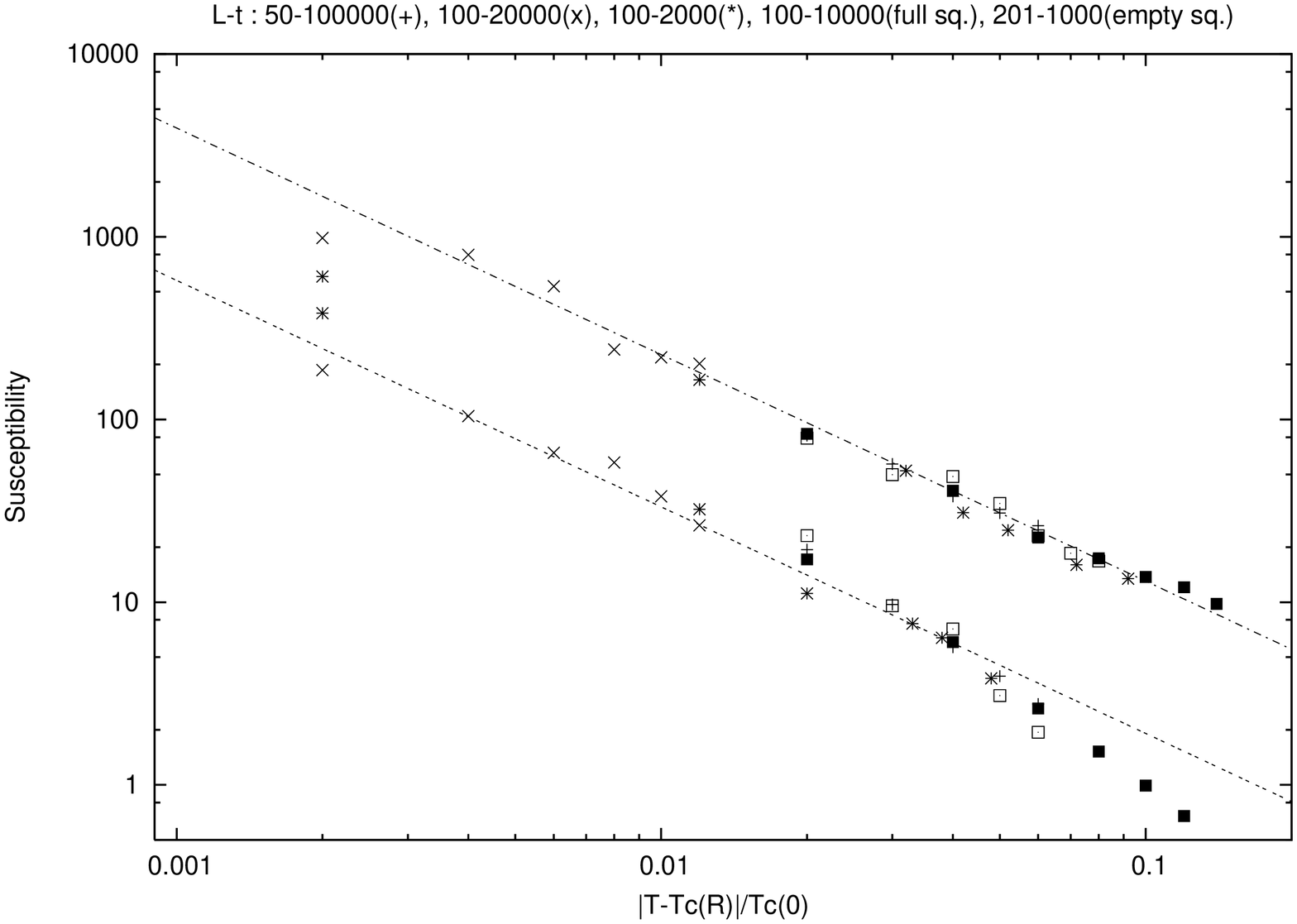}
\includegraphics[angle=-90,scale=0.31]{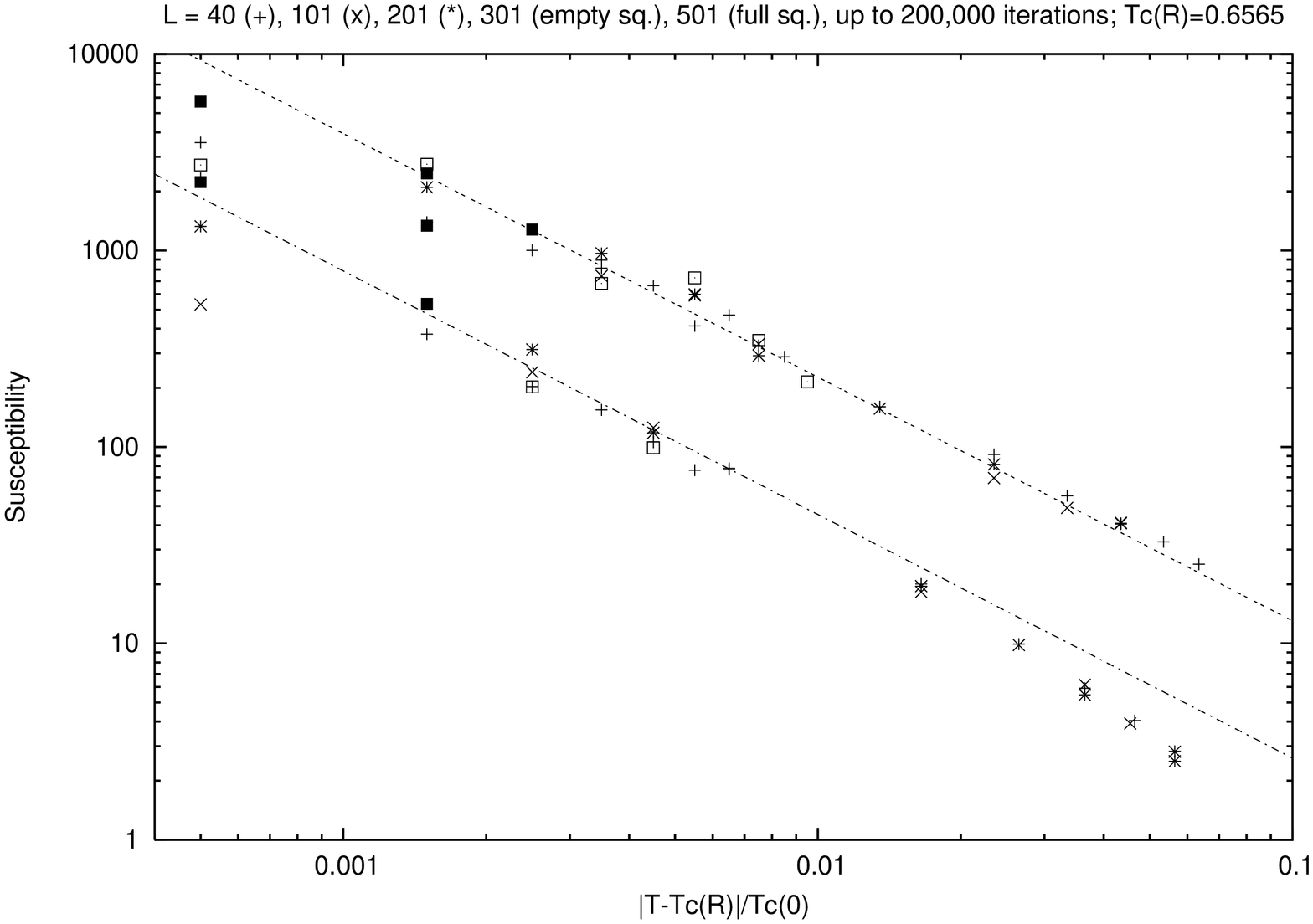}
\end{center}
\caption{As Fig.1 but with 6 NNN neighbours added to the 6 NN neighbours;
interaction ratio $R = - 0.237$. The lower part includes larger systems.
}
\end{figure}

\begin{figure}[hbt]
\begin{center}
\includegraphics[angle=-90,scale=0.5]{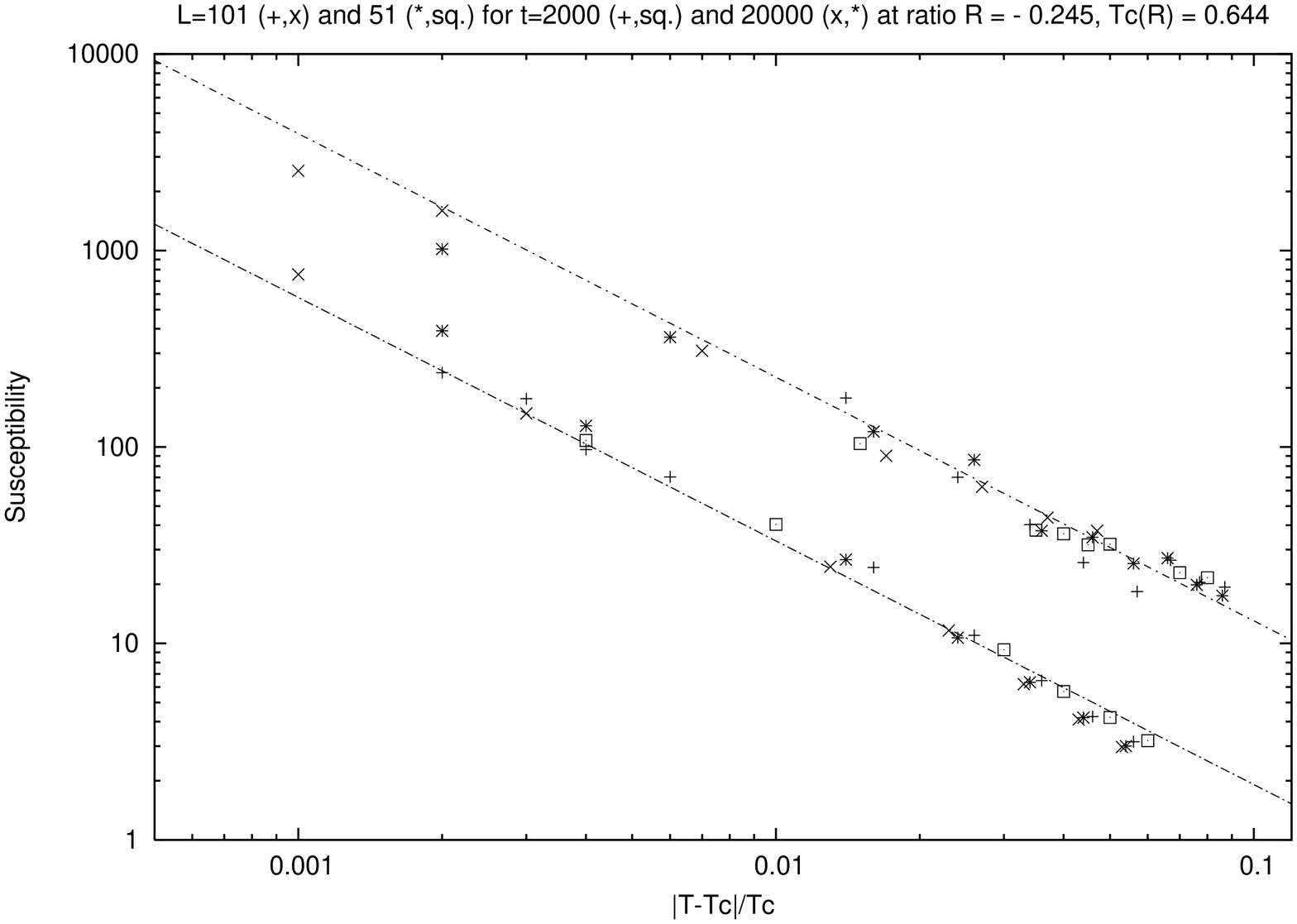}
\end{center}
\caption{As Fig.3 but for $R=-0.245$, as in [3].
}
\end{figure}

1. {\bf Introduction}

To study the critical phenomena of any system, all systems are divided [1] into 
a small number of universality classes. They are characterized by the 
dimensionality of the space and the number of components of the order parameter.
Within a certain universality class, the universal quantities (critical exponents, amplitude ratios, and scaling functions) are independent of microscopic details, such as the particular type of interactions or lattice structure.

Once the universal quantities of a universality class are known, the asymptotic critical behavior of very different systems (e.g. fluids and magnets) is believed to be known completely provided that only two nonuniversal amplitudes and
the non-universal critical temperature $T_c$ are given.

Here the 3D Ising universality class is considered by studying the susceptibility of nearest neighbour Ising model (NN model), and the next-nearest neighbour Ising model (NNN model) with only six NNN. Here some deviations from 
universality were predicted [2] and partially confirmed [3,4]. Also with 
directed interactions problems occur in the Ising model [5,6,7].
Note that Chen and Dohm [2] made no prediction on the susceptibility ratio [8]
and thus are not directly tested in the present paper, which only checks on a 
result of [3]. 

In the following section, we study the susceptibility for the NN model and the ratio of the susceptibilities above to below $T_c$, and then we simulate the
NNN model for 3D Ising model for different interaction ratios $R$, all with
Glauber kinetics. Periodic and helical boundary conditions were used throughout.
The susceptibilities were calculated through the fluctuations of the 
magnetisation while Ref.3 used the those of the absolute value of the 
magnetisation.

\bigskip
2. {\bf NN Ising Model for 3D}

To test the variation of susceptibility within the 3D Ising universality class, we choose first the NN model without external field on the simple cubic lattice
of $L \times L \times L$ spins.  Each lattice site has 6 nearest neighbours. We 
take two sizes of lattices ($L=50$ and 100), where $L$ is the size of lattice,
and then plot the susceptibility versus $|T-T_c|/T_c$ double-logarithmically
in Fig.1.  As we see there is little difference in the susceptibility 
between the two lattice sizes.

Fig.2 shows the ratio of susceptibilities above and below $T_c$ for both 
lattices, $L=50$ and 100, at the same distance from $T_c$, versus this distance.
It is consistent with the well established value near 5, and thus confirms
our simulation and analysis methods.

\bigskip
3. {\bf NNN Ising Model for 3D}

Also in the case of 6 NN and 12 NNN interactions of equal strength, we find a
susceptibility ratio near 5 (not shown). 
Furthermore we study the universality of susceptibilities in the zero field NNN 
Ising model with antiferromagnetic anisotropic next-nearest neighbour coupling,
added to the ferromagnetic NN coupling with a negative ratio $R$ of NNN to
NN interaction strength.  The anisotropic NNN Ising model [2] is established by considering only 6 of the 12 next-nearest neighbours being effective for NNN interaction, and the other 6 NNN having no interaction. The interacting NNN have the
position differences $\pm (1,1,0),\; \pm (1,0,1), \; \pm(0,1,1)$ on a simple cubic lattice.
And we do simulations for different sizes of lattices, times and temperatures,
and same coupling ratios $R =-0.237$ and $-0.245$ as used as the extreme cases
in [3]. 
Sizes between 40 and 501 and times between 1000 and 200,000 are given in the
headlines of Fig.3. The critical temperatures were determined by the maxima of 
the susceptibility, like $0.656_5$ at R + -0.237, in units of $T_c$ for the NN 
model.

Fig.3 shows the susceptibility versus the temperature difference to $T_c(R)$ 
as a log-log plot. Since the resulting susceptibility ratio is consistent with 
the one shown for only nearest neighbours and not consistent with the increased
value found by Schulte and Drope [3], we repeated the simulations at a different
place using a different computer and larger lattices up to $L = 501$; the
results in Figs.3 and 4 confirm the universality of the susceptibility ratio.

In general one lattice only was simulated, and the errors thus can best be seen
by comparing our different symbols at the same temperature, corresponding 
to different times and different size; then one also sees the systematic
errors. Fig.3b shows that the data closest to $T_c$ are too much influenced
by these systematic errors, while the two upper full squares in that Fig.3b 
at a relative temperature difference of 0.0015 correspond to $L = 501$ with
100,000 (bad) and 200,000 (better) iterations. The critical temperature was
determined from the maximum of the susceptibility. 

Ref.3 only simulated temperatures 3 \% above and below $T_c$ which is in the 
asymptotic regime for the nearest-neighbour case Fig.1 but not for the more
complicated lattice below $T_c$ in Fig.3. Perhaps that is the reason why they
got too high susceptibility ratios; our data using a wide range of temperature
differences show that 1 \% difference would have been better and that at this 
smaller difference no reliable deviation from a universal ratio exists.

\bigskip
{\bf 4. Conclusion}

The susceptibility ratio, as we see from our simulations for different $ R=0, \;
-0.237,\; -0.245$, keeps the universality of the 3D Ising model above and below 
the critical temperature. For 6 nearest neighbours, for 6+12 nearest next-nearest
neighbours and for 6+6 nearest neighbours, there is no reliable difference for the
susceptibility ratio, contradicting [3] but compatible with the standard ratio
of $4.7_5$ [9]. We did not attempt to simulate the 
more subtle universality questions connected with Binder cumulants [2,4].

We thank W. Selke for comments on the manuscript.
DS thanks Universidade Federal Fluminense in Brazil for its hospitality during 
the time of his simulations.

\bigskip
{\bf References}

\parindent 0pt

[1] K. G. Wilson, Phys. Repts. 12, 75 (1974); Rev. Mod. Phys. 47, 773 (1975).

[2] X. S. Chen and V. Dohm,  Phys. Rev. E 70, 056136 (2004).

[3] M. Schulte and C. Drope, Int. J. Mod. Phys. C 16, 1217 (2005).

[4] W. Selke and L.N. Shchur, J. Phys. A 39, L 739 (2005).

[5] M. A. Sumour and M. M. Shabat, Int. J. Mod. Phys. C 16, 585 (2005);

[6] M. A. Sumour, M. M. Shabat and D. Stauffer, Islamic University 
    Conference, March 2005; to be published in the university magazine.

[7] F. W. S. Lima and D. Stauffer, Physica A 359, 423 (2005).

[8] D.D. Betts, A.J. Guttmann, G.S. Joyce, J. Phys. C 4, 1994 (1971).

[9] M. Hasenbusch, Int. J. Mod. Phys. V 12, 911 (2001), eq.(158) and table 16.
\end{document}